\documentclass[pra,twocolumn,aps,groupedaddress,showpacs]{revtex4-1}
\usepackage{amsmath}
\usepackage{amssymb}
\usepackage{lineno}
\usepackage{graphicx}
\usepackage[usenames,dvipsnames]{color}
\usepackage{bm}
\usepackage{times}
\usepackage{hyperref}
\usepackage{float}
\hypersetup{
  colorlinks=true,
  citecolor=blue,
  linkcolor=blue,
  urlcolor=blue}

\begin{document}

\title{Bifurcations, stability, and mode evolution in segregated condensate 
       mixtures}
\author{Sukla Pal}
\author{Arko Roy}
\author{D. Angom}
\affiliation{Physical Research Laboratory,
              Ahmedabad-380009, Gujarat,
             India}

\begin{abstract}
We present new features of low energy Bogoliubov quasiparticle excitations of a 
two component Bose-Einstein condensate (TBEC) in quasi-2D geometry at zero
temperature using Hartree-Fock-Bogoliubov (HFB). We, in particular, consider 
the TBECs of $^{133}$Cs~-$^{87}$Rb and $^{85}$Rb~-$^{87}$Rb, and show
specific features in the low energy excitation spectrum as a function of 
the interaction strength. For $^{85}$Rb~-$^{87}$Rb TBEC, the appearance of a 
new zero energy mode is observed. Whereas for $^{133}$Cs~-$^{87}$Rb TBEC we 
report a bifurcation of the softened Kohn mode at the point of transition from 
miscible to immiscible domain. The lower energy mode, after the bifurcation,
goes soft and becomes a new Goldstone mode of the system.
\end{abstract}



\maketitle

\section{Introduction}

The study of quasi-low-dimensional systems is the key to understand intriguing 
aspects in the physics of ultracold atoms. It is often observed that the 
dynamics and the characteristic properties of the excitations change 
significantly with dimension. In this regard, the trapped two-component
Bose-Einstein condensates (TBECs) are interesting systems to explore. 
By choosing trapping parameters properly, the system can be made 
quasi-low-dimensional and manipulate the atomic interactions as well. Apart 
from these, the possibility of transition from miscible to immiscible phase 
makes the TBECs even more fascinating. At zero temperature, under 
Thomas-Fermi (TF) approximation 
the condition of phase separation in TBEC is given by the inequality 
$U_{12}^2>U_{11}U_{22}$~\cite{pu_98, timmermans_98}; 
where  $U_{11}$ and $U_{22}$ are the intraspecies and $U_{12}$ is the 
interspecies interaction strength. With immense experimental achievements, 
the realization of TBEC has been possible for several BEC 
mixtures of which $^{85}$Rb~-$^{87}$Rb ~\cite {papp_06} and 
$^{133}$Cs~-$^{87}$Rb~\cite{pilch_09} TBEC are considered as representative
examples in this present work, and examined in detail. In particular, the 
observation of phase separation in $^{85}$Rb~-$^{87}$Rb ~\cite{papp_08,tojo_10} 
and $^{87}$Rb~-$^{133}$Cs~\cite{mccarron_11} TBEC has opened new directions for 
various theoretical investigations in mixed systems. Among various
theoretical formalisms, Hartree-Fock-Bogoliubov theory with Popov
approximation (HFB-Popov) ~\cite{griffin_96,gies_04,gies_05} has been used 
satisfactorily in case of single-species BEC to account for the finite 
temperature effects in mode energies ~\cite{hutchinson_97, gies_04} of 
collective excitations. Besides these, it also possible to study the dynamical 
instabilities~\cite{gautam_10, kadokura_12} in $^{85}$Rb~-$^{87}$Rb TBEC at 
zero temperature. Being a gapless theory, HFB-Popov is of special interest in 
studying excitation spectrum and till now has gained much attention in 
illustrating the physics of collective excitations. Recently, the study has 
been extended for cigar-shaped TBEC in miscible ~\cite{pires_08} and in 
phase-separated domain ~\cite{roy_14a,roy_14b,roy_15}. Also, in quasi-2D 
condensate of $^{23}$Na, this theory has been successfully
applied ~\cite{roy_16} to show the modification in the excitation spectrum
due to transformation of trapping geometry at zero temperature. Therefore,
at this point it is natural to explore the new features associated with the  
mode evolution spectrum in quasi-2D TBEC. Keeping this aim in mind, in this
article, we apply HFB-Popov approximation and study systematically the 
evolution of low energy eigenmodes of Bogoliubov-de-Gennes (BdG) 
equations \cite{hu_04,japha_11} in $^{85}$Rb~-$^{87}$Rb and 
$^{133}$Cs~-$^{87}$Rb TBEC to obtain the change in quasiparticle excitation 
energy of the pancake-shaped condensate at zero temperature. The BdG equations 
have already been used extensively in characterizing the excitations of TBEC 
across miscible to immiscible phase transition ~\cite{ticknor_13}, and in 
finite temperature analysis of quasi-2D single component 
dipolar gas~\cite{ticknor_12}. 

We have, in particular, chosen $^{85}$Rb~-$^{87}$Rb TBEC to show the 
special features in excitation spectrum when two species in the TBEC have
small mass difference. The fact that the background scattering length of
$^{85}$Rb is negative, the BECs of $^{85}$Rb is achievable only after tuning
the scattering length to positive values through Feshbach 
resonance ~\cite{cornish_00, roberts_98}. So, in this case, the TBEC is driven
through the miscible-immiscible transition or vice-versa by tuning the 
intraspecies interaction of $^{85}$Rb. Most importantly, as it is the species
with the lighter mass, it also leads to the observation of Rayleigh-Taylor
instability (RTI) as the scattering length of $^{85}$Rb approaches and crosses 
the scattering length of $^{87}$Rb. The ground state of this TBEC in the 
immiscible domain shows shell structured density profile, and hence reflects 
the symmetry of the trapping potential. In contrast, $^{133}$Cs~-$^{87}$Rb 
TBEC is made up of two alkali atoms with significant mass difference, and 
one has the scope to navigate the miscible-immiscible transition by changing 
either the intraspecies or interspecies interactions. In the present work 
we vary the interspecies interaction. As the interaction is increased, the 
TBEC undergoes the miscible-immiscible transition with the side-by-side 
density profiles as the preferred geometry. In both the TBECs there are 
softening of modes associated with the emergence of RTI and 
miscible-immiscible transition. However, the trends associated with the two 
phenomena are quite different. In the former case,
the modes which go soft become zero energy modes, but in the latter, the soft
modes leads to new Goldstone modes in the system.

The paper is organized as follows: In Sec.~\ref{theory} we provide a brief
description of the HFB-Popov formalism implemented for interacting
quasi-2D TBEC. We then outline the numerical procedure employed to solve
the Bogoliubov-de-Gennes equations in. The results and discussions pertaining
to $^{85}$Rb-$^{87}$Rb and $^{133}$Cs~-$^{87}$Rb condensate mixtures are
given in Sec.~\ref{results}. The evolution of the quasiparticle excitation
energies and amplitudes with variation in intraspecies scattering length
of $^{85}$Rb are presented in Sec.~\ref{Rb}. Next, we discuss the variation
in condensate density distribution and nature of mode evolution in
$^{133}$Cs~-$^{87}$Rb condensate mixture given in Sec.~\ref{cs}.
We, then, end with conclusions highlighting the key findings of the present
work in Sec.~\ref{conc}. 

\section{Theory}
\label{theory}
We consider a TBEC in an anisotropic trap with strong axial binding such that 
the frequencies of the harmonic trapping potential satisfy the condition 
$\omega_{\perp}  \ll \omega_z$ with $\omega_x=\omega_y\equiv\omega_{\perp}$,
a configuration which is also referred as pancake (disk) shaped. With this 
choice, the TBEC remains in the ground state along the axial direction, 
and degrees of freedom are frozen. The system is thus effectively in the 
quasi-two-dimensional regime where the excitations are only along the radial 
direction. Under the mean field approximation, to incorporate the effects of 
quantum fluctuations around the ground state at zero temperature, we resort to 
the second-quantized form of the grand-canonical Hamiltonian given by
\begin{eqnarray}
\hat{H}&=&\sum_{\substack{k=1,2}}\iint dxdy\hat{\Psi}_k^{\dagger}(x,y,t)
\Bigg[-\frac{\hbar^2}{2m_k}\nabla^2_{\perp}+V_k(x,y)-\mu_k 
                       \nonumber \\
       && +\frac{U_{kk}}{2}\hat{\Psi}_k^{\dagger}(x,y,t)\hat{\Psi}_k(x,y,t)\Bigg]
\hat{\Psi}(x,y,t) +~~U_{12}\iint dxdy
                       \nonumber \\
       && \times \hat{\Psi}_1^{\dagger}(x,y,t)
\hat{\Psi}_2^{\dagger}(x,y,t)\hat{\Psi}_1(x,y,t)\hat{\Psi}_2(x,y,t),
\label{ham}
\end{eqnarray}
with $k=1,2$ as the species label, $\hat{\Psi}_k$ ($\hat{\Psi}_k^{\dagger}$) 
are the Bose field annihilation (creation) operators of the two species, and 
$\mu_k$s are the chemical potentials. The strength of the coupling constants 
are given by $U_{kk} = 2a_{kk}\sqrt{2\pi\lambda}$ and 
$U_{12}=2a_{12}\sqrt{2\pi\lambda}(1+ m_1/m_2$. It is to be 
noted that for pancake shaped traps the anisotropy parameter, as mentioned
earlier, $\lambda =\omega_z/\omega_{\perp} \gg 1$, and the form of the confining
potential is  $V(x,y,z)=(1/2)m\omega_{\perp}^2 (x^2+y^2+\lambda^2z^2)$. In 
the present work we consider $a_{kk}$ and $a_{12}$, the intraspecies and 
interspecies scattering lengths, respectively, to be positive (repulsive). 
In TBECs, an important phenomenon is the transition from miscible to immiscible 
phases or vice-versa. The miscible-immiscible transition is governed by the 
strength of the inter- and intraspecies interaction. Under the Thomas-Fermi 
approximation, the immiscible phase is characterized by the condition 
$U_{12}^2 > U_{11}U_{22}$ \cite{ho_96,pu_98,timmermans_98}. 
This inequality holds true when all the interactions in the TBEC are repulsive.

 As the ground state is macroscopically occupied for the temperature regime 
pertinent to the experiments, the condensate part can be separated out from the 
Bose field operator through the Bogoliubov decomposition 
$\hat{\Psi}_k = \phi_k + \tilde{\psi}_k$, where
$\phi_k=\langle\hat{\Psi}_k\rangle$s are the $c$-fields representing each
of the condensate species, and $\tilde{\psi}_k$s are the corresponding 
non-condensate densities or fluctuations. These fluctuations may be either
be quantum or thermal. Furthermore, using HFB-Popov approximation 
\cite{griffin_96}, $\phi_k$s are obtained as the stationary solutions of the 
coupled generalized Gross-Pitaevskii equations
\begin{eqnarray}
 \hat{h}_k\phi_k + U_{kk}\left[n_{ck}+2\tilde{n}_{k}\right]\phi_k
  +U_{12}n_{3-k}\phi_k=0,
\label{gpem}
\end{eqnarray}
where, $n_{ck}(x,y) \equiv |\phi_k(x,y)|^2$, $\tilde{n}_k(x,y) 
\equiv \langle\tilde{\psi}^{\dagger}(x,y,t)\tilde{\psi}(x,y,t)\rangle$
and $n_k(x,y)=n_{ck}(x,y)+\tilde{n}_k(x,y)$ represent the density of local
condensate, non-condensate and total condensate respectively;
$\hat{h}_{k} = (-\hbar^{2}/2m_k)\nabla_{\perp}^2 + V_k(x,y)-\mu_k$ is the
one-body part of the grand canonical Hamiltonian presented in Eq.~(\ref{ham}).
In terms of quasiparticle modes the fluctuations represented by 
$\tilde{\psi}(x,y,t)$ are of the form
\begin{eqnarray}\label{bgtrans}
\begin{aligned}
\tilde{\psi}_k=\sum_{j}\big[u_{kj}(x,y)\hat{\alpha}_j(x,y)
e^{-iE_jt/\hbar}\\- v_{kj}^{\ast}(x,y)
\hat{\alpha}_j^{\dagger}(x,y)e^{iE_jt/\hbar}\big]
\end{aligned}
\end{eqnarray}
where, $\hat{\alpha}_j$ ($\hat{\alpha}_j^{\dagger}$) are the quasiparticle 
annihilation (creation) operator satisfying Bose commutation relations, and is 
considered common to both the species. The subscript $j$ represents the energy 
eigenvalue index, and the functions $u_{jk}$ and $v_{jk}$ are the Bogoliubov 
quasiparticle amplitudes corresponding to $j$th energy eigen-state. The 
quasiparticle amplitudes are normalized as
\begin{equation}\label{norm}
\iint dxdy\sum_k\left(|u_k(x,y)|^2-|v_k(x,y)|^2\right)=1.
\end{equation}
Using the above definitions and considerations, the Bogoliubov-de-Gennes 
equations for a TBEC system are  \cite{ticknor_13,roy_14a}
\begin{subequations}
\begin{eqnarray}
 \hat{{\mathcal L}}_{1}u_{1j}-U_{11}\phi_{1}^{2}v_{1j}+U_{12}\phi_1 \left 
   (\phi_2^{*}u_{2j} -\phi_2v_{2j}\right )&=& E_{j}u_{1j},\;\;\;\;\;\;\\
    \hat{\underline{\mathcal L}}_{1}v_{1j}+U_{11}\phi_{1}^{*2}u_{1j}-U_{12}
    \phi_1^*\left (\phi_2v_{2j}-\phi_2^*u_{2j} \right ) 
     &=& E_{j}v_{1j},\;\;\;\;\;\;\\
    \hat{{\mathcal L}}_{2}u_{2j}-U_{22}\phi_{2}^{2}v_{2j}+U_{12}\phi_2\left 
    ( \phi_1^*u_{1j}-\phi_1v_{1j} \right ) &=& E_{j}u_{2j},\;\;\;\;\;\;\\
\hat{\underline{\mathcal L}}_{2}v_{2j}+U_{22}\phi_{2}^{*2}u_{2j}-U_{12} 
\phi_2^*\left ( \phi_1v_{1j}-\phi_1^*u_{1j}\right ) &=& 
E_{j}v_{2j},\;\;\;\;\;\;\;\;\;
\end{eqnarray}
\label{bdg2m}
\end{subequations}
where $\hat{{\mathcal L}}_{1}=\big(\hat{h}_1+2U_{11}n_{1}+U_{12}n_{2})$, 
$\hat{{\mathcal L}}_{2}=\big(\hat{h}_2+2U_{22}n_{2}+U_{12}n_{1}\big)$ 
and $\hat{\underline{\cal L}}_k  = -\hat{\cal L}_k$. To solve the above 
eigenvalue equations, $u_{kj}$s and $v_{kj}$s are decomposed into a linear 
combination of harmonic oscillator eigenstates followed by the diagonalisation 
of the Bogoliubov-de-Gennes matrix (BdGM) constructed from Eq.~\ref{bdg2m}. 
The order parameters $\phi_k$s and the non-condensate densities $\tilde{n}_k$s 
are then the self-consistent solutions of the coupled Eqns.~(\ref{gpem}) 
and (\ref{bdg2m}). The thermal components, in terms of the quasiparticle 
amplitudes, are defined to be 
\begin{equation}
 \tilde{n}_k=\sum_j\left [(|u_{kj}|^2+|v_{kj}|^2)N_{0}(E_j)+|v_{kj}|^2\right ],
 \label{n_k2}
\end{equation}
where, $N_0(E_j) = (e^{\beta E_j} - 1) ^{-1}$ with $\beta=1/(k_{\rm B}T) $ is 
the Bose factor of the $j$th quasiparticle mode at temperature $T$. In the 
above expression the term $|v_{kj}|^2$, independent of $N_0(E_j)$ and hence, 
the temperature, represents the quantum fluctuations. As $T$ approaches 
zero, the role of thermal fluctuations diminishes, and at $T = 0$ the 
contribution from thermal fluctuations ceases completely 
since $N_{0}(E_j)$ in Eq.(\ref{n_k2}) vanishes. The non-condensate density is 
then governed by quantum fluctuations only as Eq.~(\ref{n_k2}) reduces 
to $\tilde{n}_k=\sum_{j}|v_{kj}|^2$. Thus at 
finite temperatures the non-condensate density has dominant contribution from 
thermal fluctuations as well as quantum fluctuations.  

   To obtain the Bogoliubov quasiparticle amplitudes, we adopt following 
numerical scheme. At first, we numerically solve the pair of coupled GP
Eqns.~(\ref{gpem}) using the split step Crank-Nicholson (CN) method.
Using these solutions, the BdG Eqns.~(\ref{bdg2m}) are then cast as a
matrix eigenvalue equation in the basis of the harmonic oscillator
potential eigenstates. Then we write $u_k$ and $v_k$'s as linear combination
of the harmonic oscillator direct product states 
$\varphi(x) \otimes \varphi(y)$, where $\varphi(x)$ and $\varphi(y)$ are the 
harmonic oscillator eigenstates in $x$ and $y$ direction, respectively. 
With this definition,  
\begin{equation}
 u_{1j}(x,y) = \sum_{\substack{k,l=0}}^{N_b}p_{jkl}\varphi_{kj}(x)
 \varphi_{lj}(y)
\end{equation}
where, $p_{jkl}$ is the coefficients of linear combination. Similarly, we can 
define $u_{2j}$, $v_{1j}$, and $v_{2j}$ as linear combinations of the direct
product states. Using the above definition, for equal number of basis functions
$N_b$ along  the $x$ and $y$ axis, the BdG matrix is of dimension
$4(N_b+1)(N_b+1)\times 4(N_b+1)(N_b+1)$. Considering the orthogonality of
harmonic oscillator basis, the resulting BdG matrix is a sparse
matrix. Due to the  $N_b^2$ scaling of the BdG matrix, the matrix size rapidly 
increases with the basis size, and it is essential to use algorithms capable
of large matrix diagonalization. For this reason we use
ARPACK \cite{lehoucq_98} routines to diagonalises the BdG matrix, and 
consequently, we consider a selected set of the quasiparticle amplitudes in the 
computation of fluctuations or non-condensate density. This is done such that, 
only the very high energy modes and hence, negligible Bose factor, 
quasiparticle amplitudes are excluded from the computation of non-condensate 
density. The non-condensate density is computed using the Eq.~(\ref{n_k2}), 
and we iterate the 
solutions until the condensate, and non-condensate densities converge to  the 
predefined accuracies. To accelerate the convergence we use the method of 
successive under-relaxation, and choose the under-relaxation parameter 
$S = 0.1$ \cite{simula_01}. The new solution at the $i$th iteration is then
\begin{equation}
\phi_i^{new}(x,y)= S\phi_i(x,y) + (1-S)\phi_{i-1}(x,y)
\label{relax}
\end{equation}
where $i$ is the iteration index. Since we focus on zero temperature 
excitations, the low energy eigenmodes will be sufficient to take care of 
the quasiparticle amplitudes.
\begin{figure}[t]
 \includegraphics[width=8.0cm]{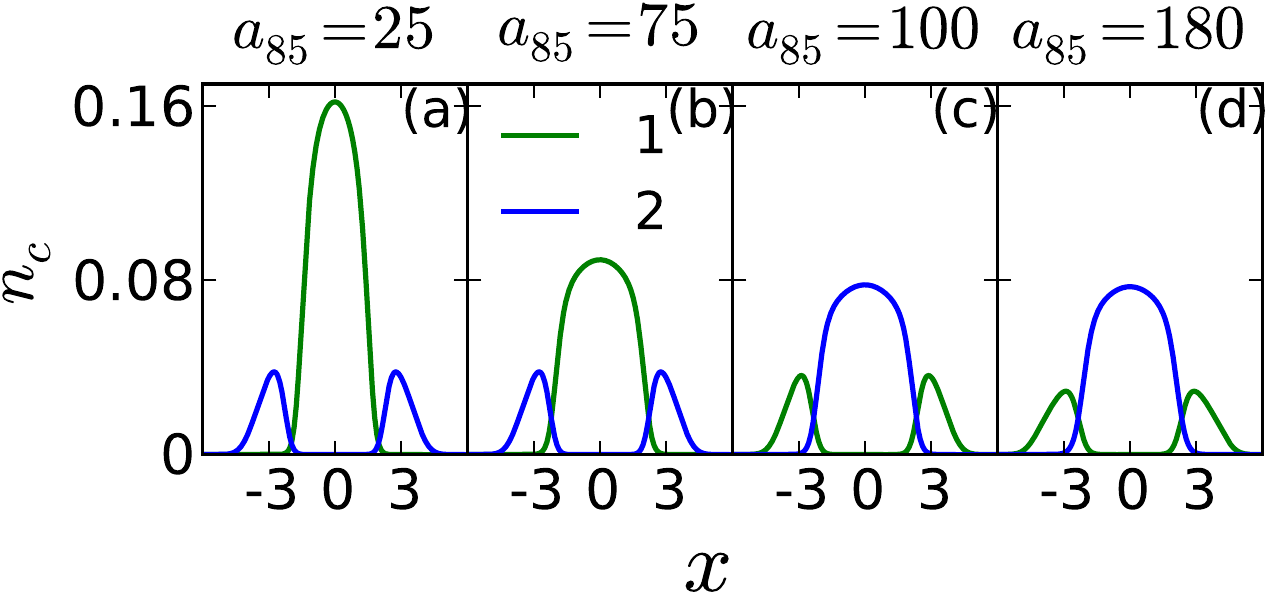}
    \caption{(Color online)Equilibrium density profile of $^{85}$Rb-$^{87}$Rb 
             along $x$-axis with $y = 0$ for 
             $a_{\rm ^{85}Rb}$ = 25, 75, 100, 180 $a_0$ at zero temperature. 
             The density is measured in units of $a^{-2}_{\rm osc}$. 
             (a)-(b) $^{85}$Rb condensate occupies the central region of the 
             trap, and $^{87}$Rb condensate is at the edges. 
             (c)-(d) As $a_{85}$ 
             is increased when $a_{\rm ^{85}Rb} > a_{\rm ^{87}Rb}$, the 
             density profiles switch their positions. The $x$ and $y$ are 
             measured in units of $a_{\rm osc}$.}
    \label{rbrb-den}
\end{figure}


\section{Results and Discussions}
\label{results}
The low-lying quasiparticle spectrum of a trapped quasi-2D TBEC is
characterized by the presence of two Goldstone modes which appear due to
breaking of $U(1)$ global gauge symmetry for each of the condensate species, 
and the Kohn modes~\cite{kohn_61, brey_89, dobson_94, fetter_98}. The Kohn or 
dipole modes, among the low-lying energy eigenmodes have maximum $N_0$, and 
have the dominant contributions to the quantum fluctuations and thermal 
densities. For this reason, we investigate and examine the evolution of these 
modes, and in particular, study the topological deformation of the
quasiparticle amplitude corresponding to the Kohn mode for various phases
of the TBEC.

In the present study, we examine the evolution of the quasiparticle modes
in TBEC systems with the variation in interaction strengths which drives the
system from miscible to immiscible regime or vice-versa. The variation can
either be the intra or the interspecies scattering length of
the atoms constituting the TBEC. An representative example of the first
possibility,  tuning intraspecies interaction, is the TBEC of  
$^{85}$Rb~-$^{87}$Rb, where the intraspecies scattering length of $^{85}$Rb can 
be tuned experimentally via a magnetic Feshbach resonance~\cite{papp_08}. It 
must 
be mentioned here that, it is essential to tune the intraspecies scattering 
length of $^{85}$Rb as it has negative background scattering length. In this 
mixture, as the atomic masses are nearly same, the energetically favorable
ground state configuration is the species with the lower
repulsive interaction strength being surrounded by the species with higher
repulsive interaction strength. The representative example of the other case,
tuning the interspecies scattering length, is the TBEC 
of $^{133}$Cs~-$^{87}$Rb,
where the interspecies scattering length can be tuned through magnetic
Feshbach resonance~\cite{pilch_09,mccarron_11}. In this case the 
stable ground state configuration is Cs atoms being surrounded by Rb atoms. We 
choose these representative systems and theoretically investigate the nature of 
excitation spectra in the miscible and immiscible phases of these systems at 
$T=0$. Based on our previous works, the qualitative features of the results 
from these two examples are applicable to the TBECs of other possible atomic 
species.
\begin{figure}[t]
 \includegraphics[width=8.5cm]{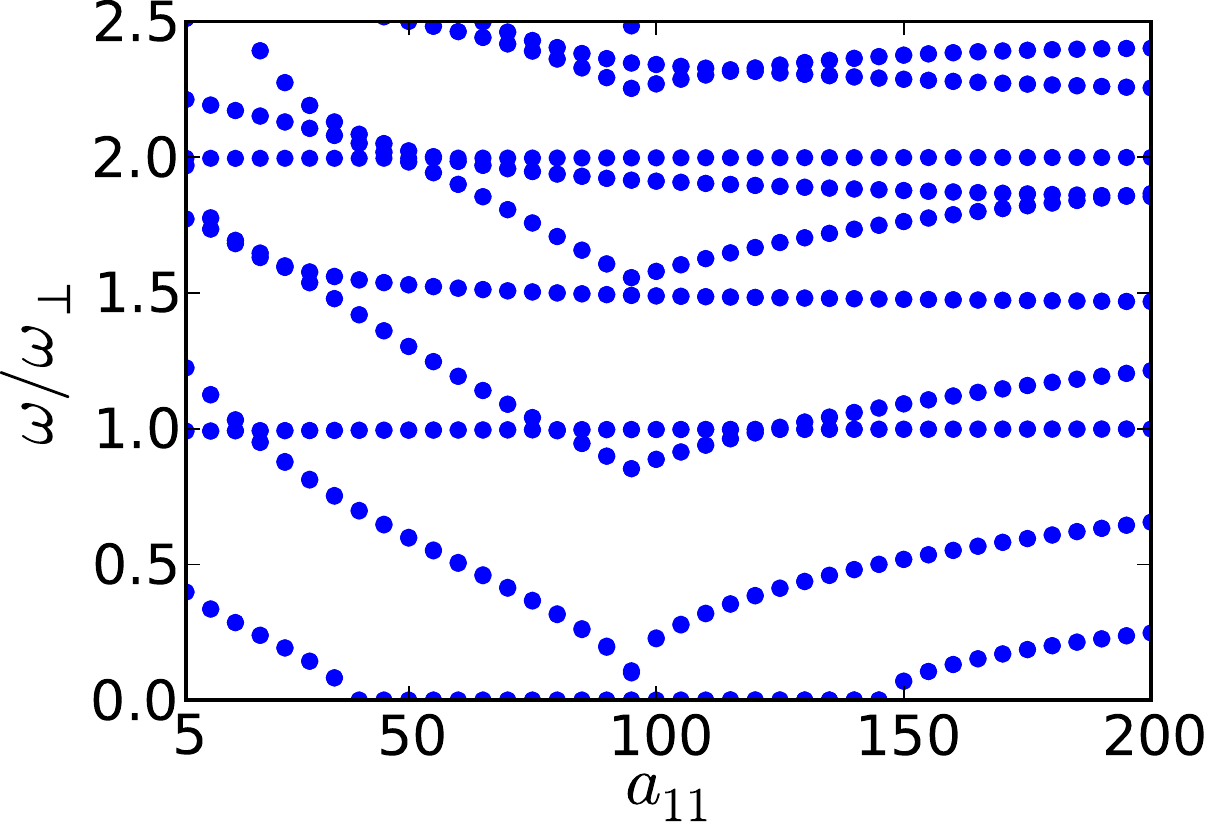}
    \caption{(Color online) The evolution of low-lying mode energies as a 
             function of $a_{11}$ in the domain $5a_0\leqslant
             a_{11}\leqslant 200a_0$ at zero temperature for 
             $N_1 = N_2 = 5000$. Here $a_{11}$ is in units of $a_0$.}
 \label{rb-mode-evolve}
\end{figure}


\subsection{Mode evolution of $^{85}$Rb~-$^{87}$Rb BEC mixture}
\label{Rb}
To examine the quasiparticle excitation spectrum of $^{85}$Rb~-$^{87}$Rb TBEC 
with our theoretical scheme, we consider $^{85}$Rb and $^{87}$Rb as species 
1 and 2, respectively. The interspecies scattering length is 
$a_{12}=a_{\rm ^{85}Rb-^{87}Rb}=214a_0$, where $a_0$ is the Bohr radius. 
Similarly, the intraspecies scattering lengths are denoted by 
$a_{22}=a_{\rm ^{87}Rb}$, and $a_{11}= a_{\rm 85}$. In the present study, 
as mentioned earlier, we change $a_{11}$ while keeping $a_{22}$ fixed 
at $99a_0$. We consider equal number of particles for both the species 
$N_{\rm ^{85}Rb } = N_{\rm ^{87}Rb} = 5\times10^3$, which maybe lower than 
number of atoms in experimentally realized TBECs, but this does not affect 
the qualitative nature of the present results. To form a quasi-2D trap we set 
$\lambda = 12.5$ and $\omega_x=\omega_y=\omega_{\perp}=2\pi\times 8.0$ Hz 
\cite{neely_10}. At zero temperature, for these set of parameters, at low 
values of $a_{11}$ ($a_{11}$<$a_{22}$) the $^{85}$Rb~-$^{87}$Rb TBEC is in an 
immiscible phase with shell structured density profiles. In the 
domain $a_{11} < a_{22}$, the $^{85}$Rb condensate lies at the center of the 
trap with $^{87}$Rb condensate lying at the edges, and the positions get 
interchanged when $a_{11} > a_{22}$.
\begin{figure}[t]
 \includegraphics[width=8.5cm]{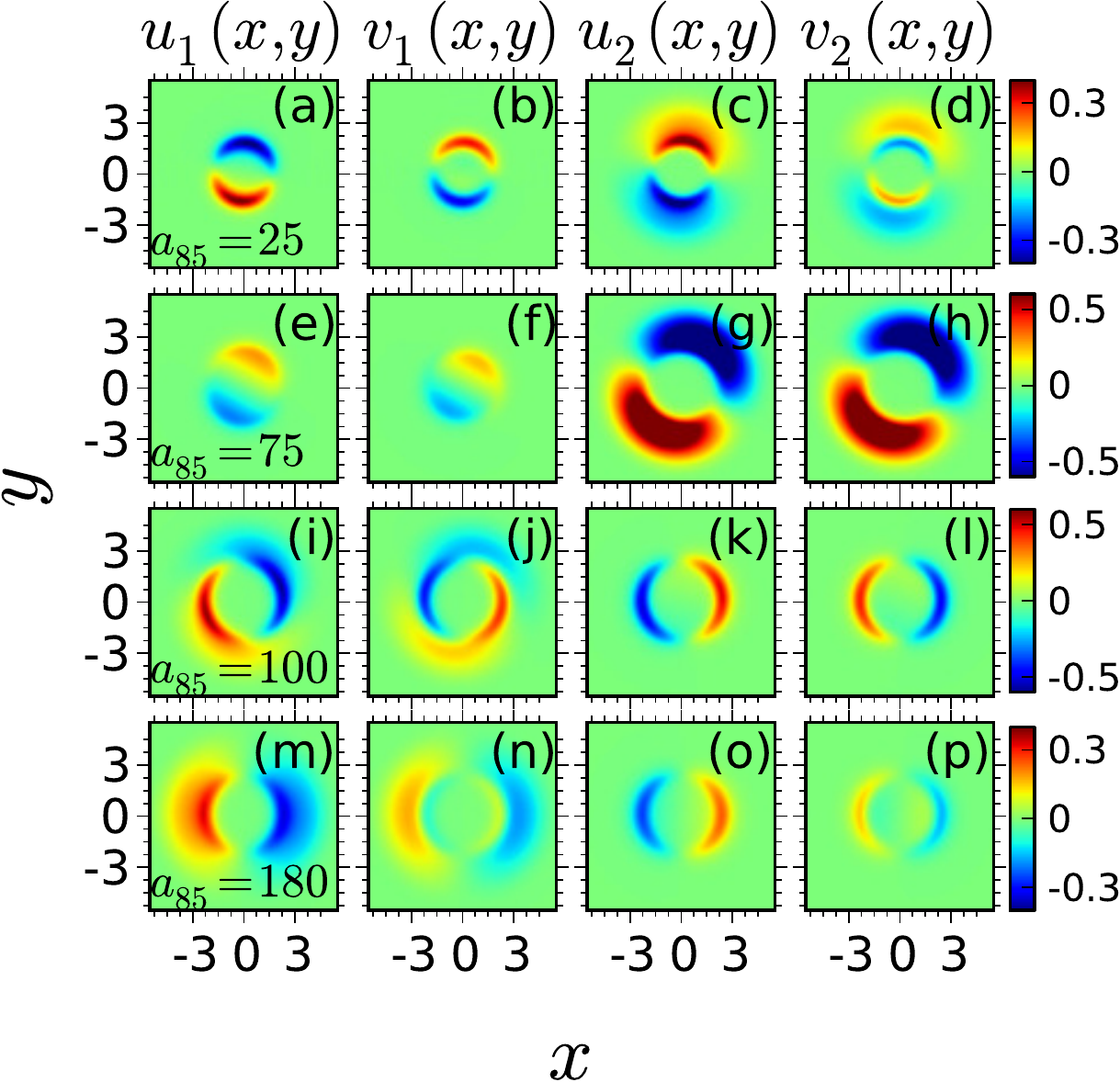}
    \caption{(Color online) The Kohn mode amplitude at selected values of 
             $a_{\rm ^{85}Rb}$, and the other scattering lengths are kept fixed.
             The chosen values of $a_{\rm ^{85}Rb}$ are representative of the 
             stages in the exchange of position between the two species 
             associated with RTI. The components of the amplitude are 
             $u_1=u_{\rm ^{85}Rb}$, $v_1=v_{\rm ^{85}Rb}$, 
             $u_2=u_{\rm ^{87}Rb}$ and $v_2=v_{\rm ^{87}Rb}$. The figures
             correspond to (a)-(d) $a_{85}$=25$a_0$, (e)-(h)$a_{85}$=75$a_0$, 
             (i)-(l)$a_{85}$=100$a_0$ and (m)-(p) $a_{85}$=180$a_0$.
             The values of $u$ and $v$ are in units of $a^{-1}_{\rm osc}$, and 
             $x$ and $y$ are measured in units of $a_{\rm osc}$.}
\label{zero-mode}
\end{figure}

In Fig.~\ref{rbrb-den} we show the  condensate density profiles
along $x$-direction as $a_{11}$ is varied from 25$a_0$ to 180$a_0$. Since 
the condensate densities follow the azimuthal symmetry of the trapping potential
in the $x-y$ plane, the corresponding density profiles along $y$ direction is 
identical to the density along $x$-axis. In the figure, 
Fig.~\ref{rbrb-den}(a)-(b) $a_{11}<a_{22}$, $^{87}$Rb surrounds $^{85}$Rb, and 
in Fig.~\ref{rbrb-den}(c)-(d), we observe the reverse configuration as 
$a_{11}>a_{22}$. These are the energetically favorable density configurations.
Furthermore significant changes in density profiles in 
Fig.~\ref{rbrb-den}(a)-(b) are worth notable with the increase of $a_{11}$.
As $a_{11}$ is increased, an important phenomenon occurs when 
$a_{11}$ is larger than $a_{22}$. Around this point Rayleigh Taylor 
instability (RTI) sets in as the two condensates swap their positions. This 
is reflected in the nature of the mode 
evolutions, and the presence of modes with complex eigenvalues. In the 
density profiles, as mentioned before, the interchange of the positions is 
evident when we compare the density profiles in Fig.~\ref{rbrb-den}(b) and 
Fig.~\ref{rbrb-den}(c).

In Fig.~\ref{rb-mode-evolve} we show the evolution of the mode energies as a 
function of $a_{11}$.  The excitation spectrum has two Kohn modes, one each for 
the two species. The energy of one of the Kohn modes remains constant at 
$\omega=\omega_{\perp}$ in the entire range of $a_{11}$ considered, 
validating Kohn's theorem, and provides an important consistency check for our 
computations. At the outset, when $a_{11}=5a_0$, that is, when the TBECs are 
strongly phase-segregated, the energy of the other or second Kohn mode is
$\omega=0.398\omega_{\perp}$. With the increase in $a_{11}$, the spatial 
extent of $^{85}$Rb condensate gets larger, thereby leading to a finite 
overlap between the two species. This influences the energy of the second 
Kohn mode which starts becoming soft, and eventually becomes a zero energy 
mode when $a_{11}\approx 40a_0$ as shown in Fig.~\ref{rb-mode-evolve}. The 
softened Kohn mode continues to be a zero energy mode 
till $a_{11}\approx145a_0$ after which it regains energy and hardens. The 
appearance of this additional zero energy mode in 
the region $40a_0\leqslant a_{11}\leqslant 145a_0$ is an indication of the
onset of energetic instability within the mixture, since in the region around 
$a_{11}=100a_0$, the species are expected to have RTI to minimize the total 
energy of the system. The onset of the RTI is also evident from the nature of
the Kohn mode energy. Albeit in Fig.~\ref{rb-mode-evolve} we have plotted
the real part of the Kohn mode energy, in the domain 
$40a_0\leqslant a_{11}\leqslant145a_0$ the mode energy has a small imaginary 
component, and this is a characteristic signature of an instability present in 
the system. Furthermore, our studies reveal that the quadrupole 
mode becomes soft with the increase in $a_{11}$, and  
at $a_{11}\approx99a_0$ it becomes a zero energy mode. So, 
at $a_{11}\approx99a_0$, in addition to the 
Nambu-Goldstone modes of the system, there are two more zero energy modes. It
must be mentioned that, the quadrupole mode in the initial stages of evolution 
collides with the Kohn mode of the system. After the collision, its energy 
continues to decrease  till it becomes zero energy mode, and afterwards
the energy increases. The softening of the Kohn and quadrupole modes are 
accompanied by structural changes in the mode structures, and trends in the 
mode energy evolutions are a consequence of the instability in the system. 
\begin{figure}[t]
 \includegraphics[width=8.5cm]{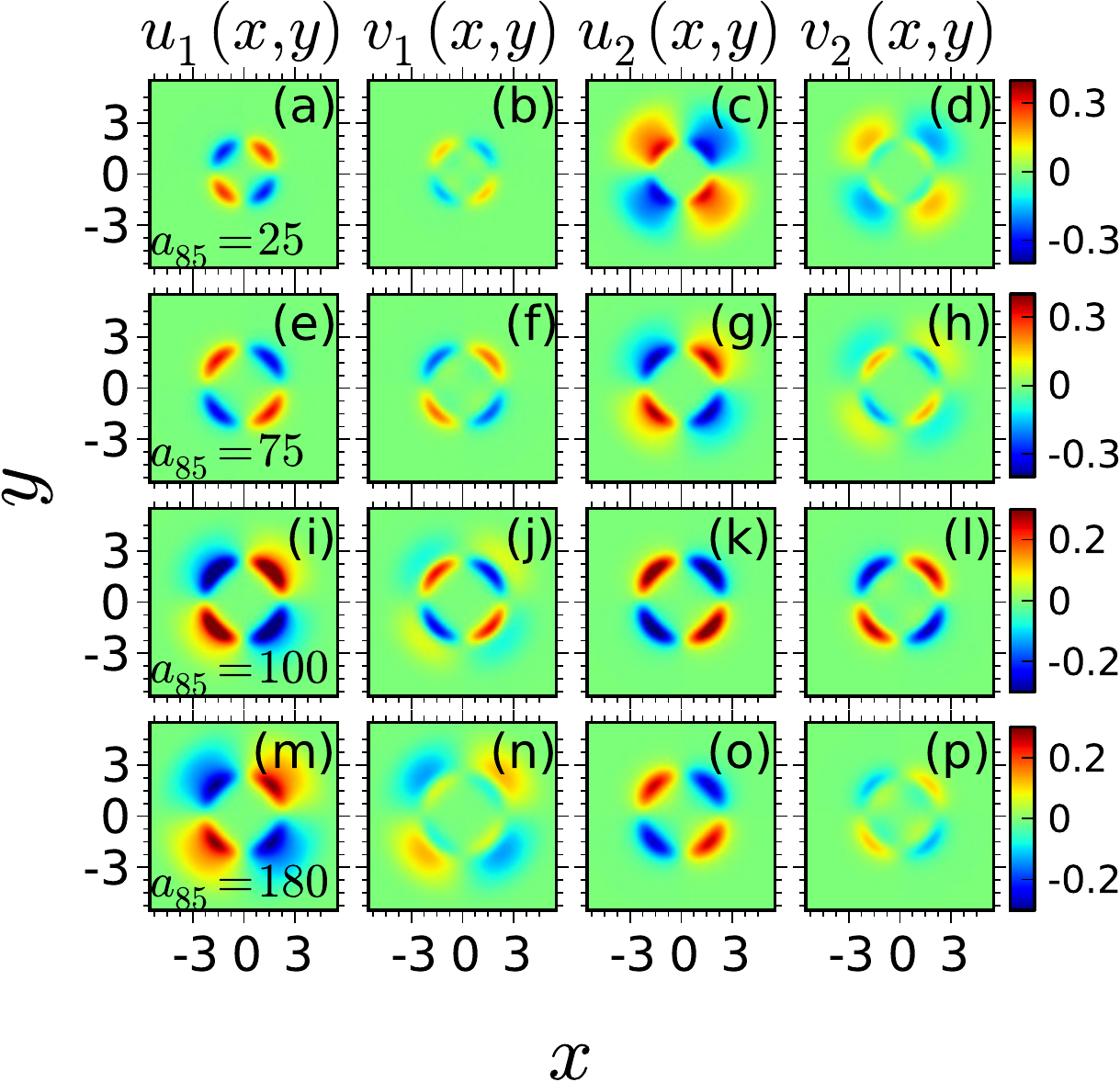}
    \caption{(Color online) The quadrupole mode amplitude at selected values of 
             $a_{\rm ^{85}Rb}$, and the other scattering lengths are kept fixed.
             The chosen values of $a_{\rm ^{85}Rb}$ are representative of the 
             stages in the exchange of position between the two species 
             associated with RTI. The components of the amplitude are 
             $u_1=u_{\rm ^{85}Rb}$, $v_1=v_{\rm ^{85}Rb}$, 
             $u_2=u_{\rm ^{87}Rb}$ and $v_2=v_{\rm ^{87}Rb}$. As $a_{^{85}Rb}$ 
             is increased the effective wavelength of $u_1$ and $v_1$ increases 
             whereas the reverse happens for $u_2$ and $v_2$. The figures
             correspond to (a)-(d) $a_{85}$=25$a_0$, (e)-(h)$a_{85}$=75$a_0$, 
             (i)-(l)$a_{85}$=100$a_0$ and (m)-(p) $a_{85}$=180$a_0$.
             The values of $u$ and $v$ are in units of $a^{-1}_{\rm osc}$, and 
             $x$ and $y$ are measured in units of $a_{\rm osc}$.
}
    \label{quad-mode}
\end{figure}

 To examine the evolution of the Kohn mode in better detail we analyse the 
structure of the quasiparticle amplitudes. For this we show the quasiparticle
amplitude for selected values of $a_{11}$ in Fig.~\ref{zero-mode}. At 
$a_{11}=25a_0$, from Fig.\ref{zero-mode} (a)-(d) it is evident that the 
mode has a ring like geometry, and dipole in structure. With the 
increase of $a_{11}$ 
as it becomes a zero energy mode amplitude, shown in Fig.\ref{zero-mode}(e)-(l),
becomes broader, and this is indicative of a decrease in the wave number of the
mode.  At $a_{11}\approx 99a_0$, the $u_1$ and $v_1$ have a mismatch of the
edges or spiral like structure, and this corresponds to the value of $a_{11}$
where the RTI is expected to occur. Finally, for $a_{11}>a_{22}$, 
quasiparticle amplitudes swap their position as is obvious from 
Fig.\ref{zero-mode}(m)-(p).

 An important observation associated with the quasiparticle amplitudes in
Fig.\ref{zero-mode} is the relative phases of the components $u_k$ and $v_k$. 
For the domain $a_{11}\approx 25a_0$, shown in Fig.\ref{zero-mode}(a)-(d),
$u_1$ is $\pi$ out of phase with $v_1$ while $u_2$ and $v_2$ are in phase. 
With increase of $a_{11}$ in the domain $a_{11}\approx 75a_0$, as shown in 
Fig.\ref{zero-mode}(e)-(h), $u_1$ and $v_1$ are in phase so are $u_2$ and 
$v_2$, however, $u_1$ ($v_1$) and $u_2$ ($v_2$) are out of phase. With further 
increase of $a_11$, at $a_{11} \approx 100$, $u_1$ is 
$\pi$ out of phase with $v_1$, and so are $u_2$ and $v_2$. At 
$a_{11}\approx 180$ when the species interchange their positions $u_1$ and 
$v_1$ are in phase while $u_2$ and $v_2$ are $\pi$ out of phase. Thus, not 
only the positions of the species, the phase difference between 
the quasiparticle amplitudes gets reversed. This restores the relative phase
differences between the species wise quasiparticle amplitudes. The different 
relative phases in the domain $a_{11}\approx 75a_0$ and $a_{11} \approx 100$ 
are the intermediate phase patterns in the transition associated with the 
exchange in
the position of the two species. A similar trend is observed in the case of
quadrupole mode as well, and are shown in Fig.\ref{quad-mode}. The variation of 
quasiparticle amplitudes corresponding to the quadrupole mode which has 
$\omega = 1.224 \omega_{\perp}$ at $a_{11} = 5a_0$ is shown in
Fig.\ref{quad-mode}. We note that as $a_{11}$ is increased the effective 
wavelength corresponding to $u_1$ and $v_1$ increases, and the reverse 
happens for the $u_2$ and $v_2$. The reason is that with the increase 
of $a_{11}$ it is 
energetically favourable for the species 1 to be at the periphery and species 2
to be at the core. As discussed earlier, the changes are also manifested in the 
mode evolution. However, the changes in the structure, and phases of the 
quasiparticle amplitudes are unique characteristics associated with the onset
of RTI.
\begin{figure}[t]
 \includegraphics[width=9.0cm]{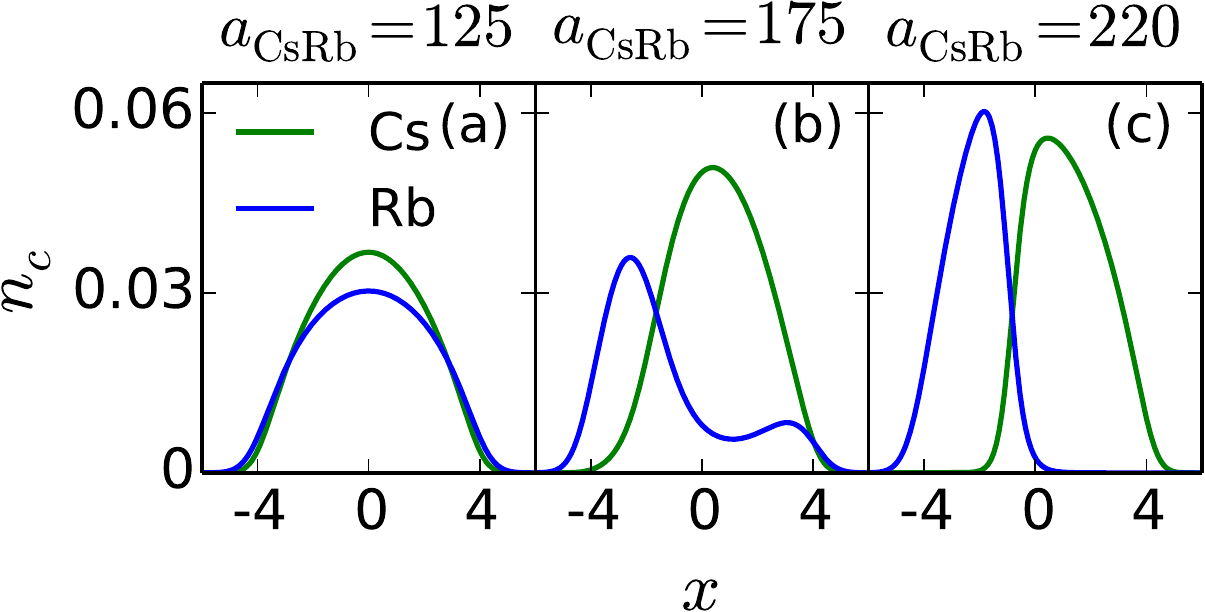}
     \caption{(Color online) Equilibrium density profiles of Cs-Rb TBEC
               along $x$-axis with $y = 0$ at $T = 0$ K showing transition
               from miscible to immiscible (phase-separated) regimes with the
               change in interspecies scattering length $a_{\rm CsRb}$.
               (a) Shows Cs-Rb TBEC in miscible phase for 
                   $a_{\rm CsRb} = 125 a_0$.
               (b) Corresponds to the Cs-Rb density profiles just on the verge
                   of phase-separation at $a_{\rm CsRb} = 175 a_0$.
               (c) Shows phase-separated density profiles of Cs-Rb TBEC for
                   $a_{\rm CsRb} = 220 a_0$. These are referred to as 
                   side-by-side density profiles. $n_c$ and $x$ are measured 
                   in units of $a_{\rm osc}^{-2}$ and $a_{\rm osc}$, 
                   respectively. 
              }
 \label{xcutCsRb}
\end{figure}


\subsection{$^{133}$Cs-$^{87}$Rb BEC mixture}
\label{cs}
A TBEC of heteronuclear atoms, and which is of current experimental interest is 
Cs-Rb TBEC~\cite{pattinson_13}. This experimental observation has been the 
stepping stone towards the realization of stable ultracold CsRb molecules in 
the rovibrational ground state~\cite{gregory_16,molony_14}. It is possible to 
steer this TBEC from the miscible to immiscible phase using interspecies 
Feshbach resonance~\cite{pilch_09}. In this system we label Cs and Rb to be 
species 1 and 2, respectively. With this identification 
$a_{11} =a_{\rm CsCs} = 280 a_0$ and $a_{22} = a_{\rm RbRb} = 100a_0$, as 
mentioned earlier $a_0$ is the Bohr radius. 
For these scattering lengths, and based on Thomas-Fermi approximation, the 
condition for phase-separation is $a_{12} = a_{\rm CsRb} > 164 a_0$, which is 
lower than 
the background $a_{\rm CsRb}\approx 650 a_0$~\cite{lercher_11}. The quasi-2D 
Cs-Rb TBEC system that we consider here corresponds to $N_1=N_2 = 2000$,
and the trapping parameters are the same as mentioned earlier. For this set
of parameters, with $a_{12} = 125 a_0$, the ground state of the system is in 
miscible phase and is rotationally symmetric. The density profile is as shown in
Fig.~\ref{xcutCsRb}(a). As $a_{12}$ is increased to higher values, at the
point of phase-separation the rotational symmetry is, however, broken at
$a_{12} \approx 175 a_0$ as shown in Fig.~\ref{xcutCsRb}(b). The Cs-Rb
condensate clouds segregate from each other at higher $a_{12}$ with minimal
interfacial overlap. They lie adjacent to each other with Cs condensate
cloud occupying one side of the trap, and Rb condensate cloud occupying the
other side. An example of the side-by-side ground-state density profile of 
Cs-Rb TBEC is shown in Fig.~\ref{xcutCsRb}(c).

\begin{figure}[H]
 \includegraphics[width=9.0cm]{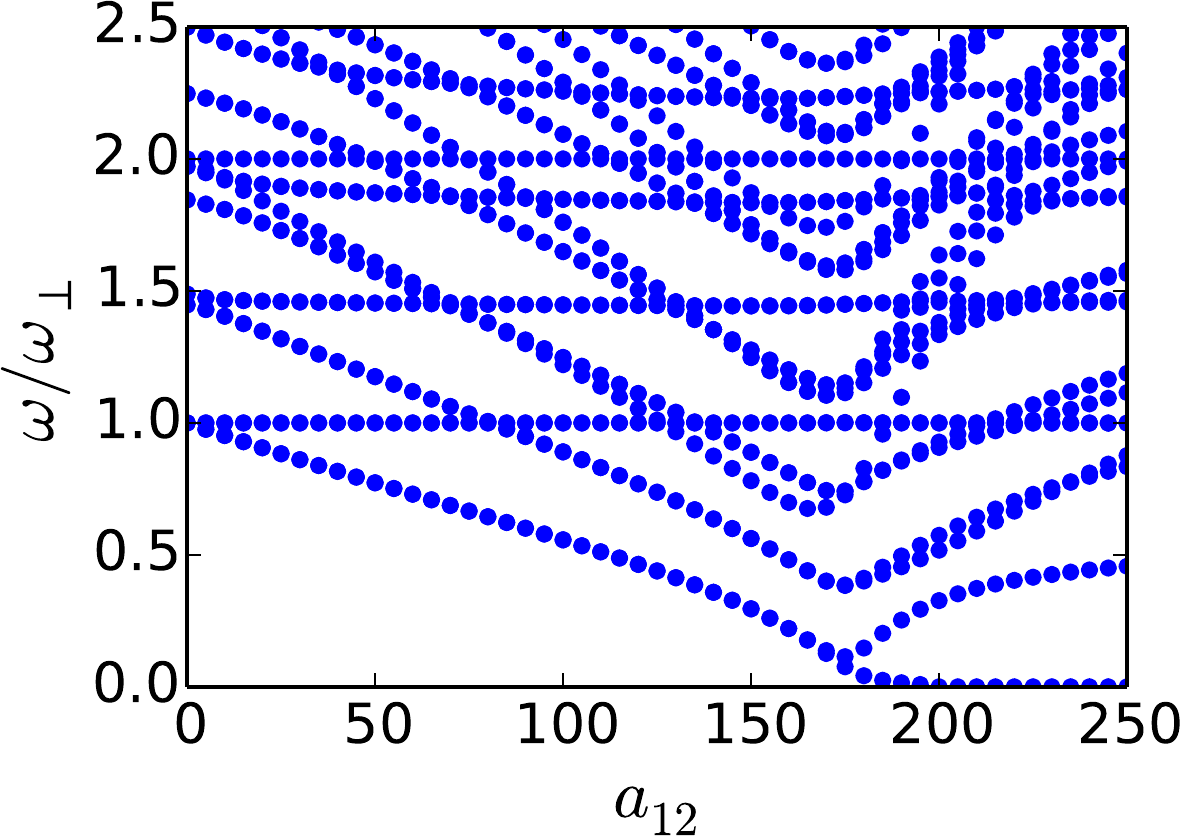}
    \caption{(Color online) The evolution of the low-lying modes in Cs-Rb TBEC
             as a function of the interspecies scattering length 
             $a_{\rm CsRb}$. Here $a_{12}$ is in units of $a_0$.
            }
    \label{mevlCsRb}
\end{figure}


\subsubsection{Mode evolution for miscible to side-by-side transition}

The low-lying excitation spectra of a quasi-2D TBEC system is described by the
presence of two Goldstone modes, and the degenerate slosh or dipole modes.
The slosh modes of the two condensates may either be out-of-phase or
in-phase. The in-phase slosh modes with center-of-mass motion are referred to 
as Kohn modes. To investigate the interaction induced modification of the 
quasiparticle spectra due to phase transition from miscible to immiscible 
regimes, we vary $a_{\rm CsRb}$ and compute the quasiparticle energies at 
zero temperature. At the outset, when $a_{\rm CsRb} = 0$, 
Eqns.~(\ref{gpem}) become decoupled and the excitation spectrum of the two 
species are independent. The slosh modes of the system then occur 
at $\omega=\omega_\perp$. These modes, however, start intermixing 
for $a_{\rm CsRb}>0$. With increasing $a_{\rm CsRb}$, 
the energy of the slosh mode decreases, but the Kohn mode remains steady
at $\omega=\omega_\perp$. At higher $a_{\rm CsRb}$, the energy of the
doubly degenerate slosh modes decreases further till it attains a minimum
value at $a_{\rm CsRb}^{\rm c} \approx 175 a_0$. The rotational symmetry of the 
condensate is then broken at $a_{\rm CsRb}^{\rm c}$, and the degeneracy of
the slosh modes is lifted, accompanied by the bifurcation of the mode energy 
into two branches. For $a_{\rm CsRb} > 175 a_0$, the Cs-Rb density profiles 
start 
segregating from each other till they become phase-separated at 
$a_{\rm CsRb} \approx 220 a_0$ as shown in Fig.~\ref{xcutCsRb}(c). With the 
variation of $a_{\rm CsRb} = 175 a_0$ to $a_{\rm CsRb} = 220 a_0$, the energy 
of the quasiparticle excitation corresponding to the lower branch 
continues to go soft and becomes a Goldstone mode. The upper branch, however, 
hardens. This trend in mode evolution is shown in Fig.~\ref{mevlCsRb}. 
\begin{figure}[H]
 \includegraphics[width=9.0cm]{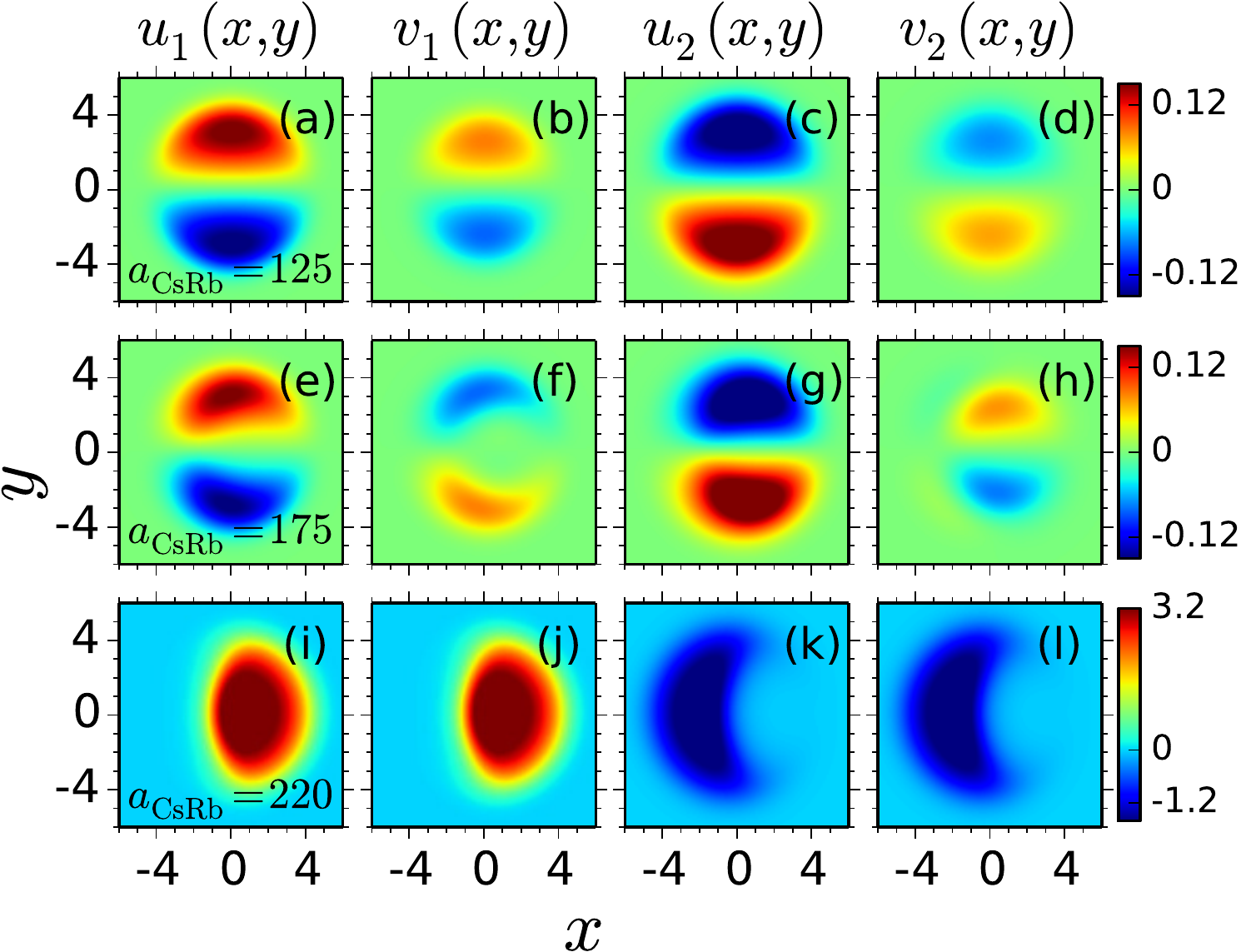}
      \caption{(Color online) Evolution of the quasiparticle amplitudes 
                corresponding to the slosh modes whose energy 
                decreases as $a_{\rm CsRb}$ is varied from
                $0a_0$ to $250a_0$. (a-d) show the quasiparticle amplitudes
                corresponding to one of the degenerate Kohn modes at 
                $a_{\rm CsRb} = 125a_0$. At $a_{\rm CsRb} = 175a_0$, that is, 
                at the point of phase-separation, the degeneracy of the Kohn 
                mode is lifted and a bifurcation in the mode evolution is 
                evident. The quasiparticle amplitudes corresponding 
                to the lower branch of the slosh mode are shown in 
                (e-h) for $a_{\rm CsRb} = 175a_0$, and 
                (i-l) for $a_{\rm CsRb} = 220a_0$.
                Here $u$s and $v$s are in units of 
                $a_{\rm osc}^{-1}$. $x$ and $y$ are measured in units of 
                $a_{\rm osc}$.
              }
      \label{qampCsRb}
\end{figure}

\begin{figure}[H]
 \includegraphics[width=9.0cm]{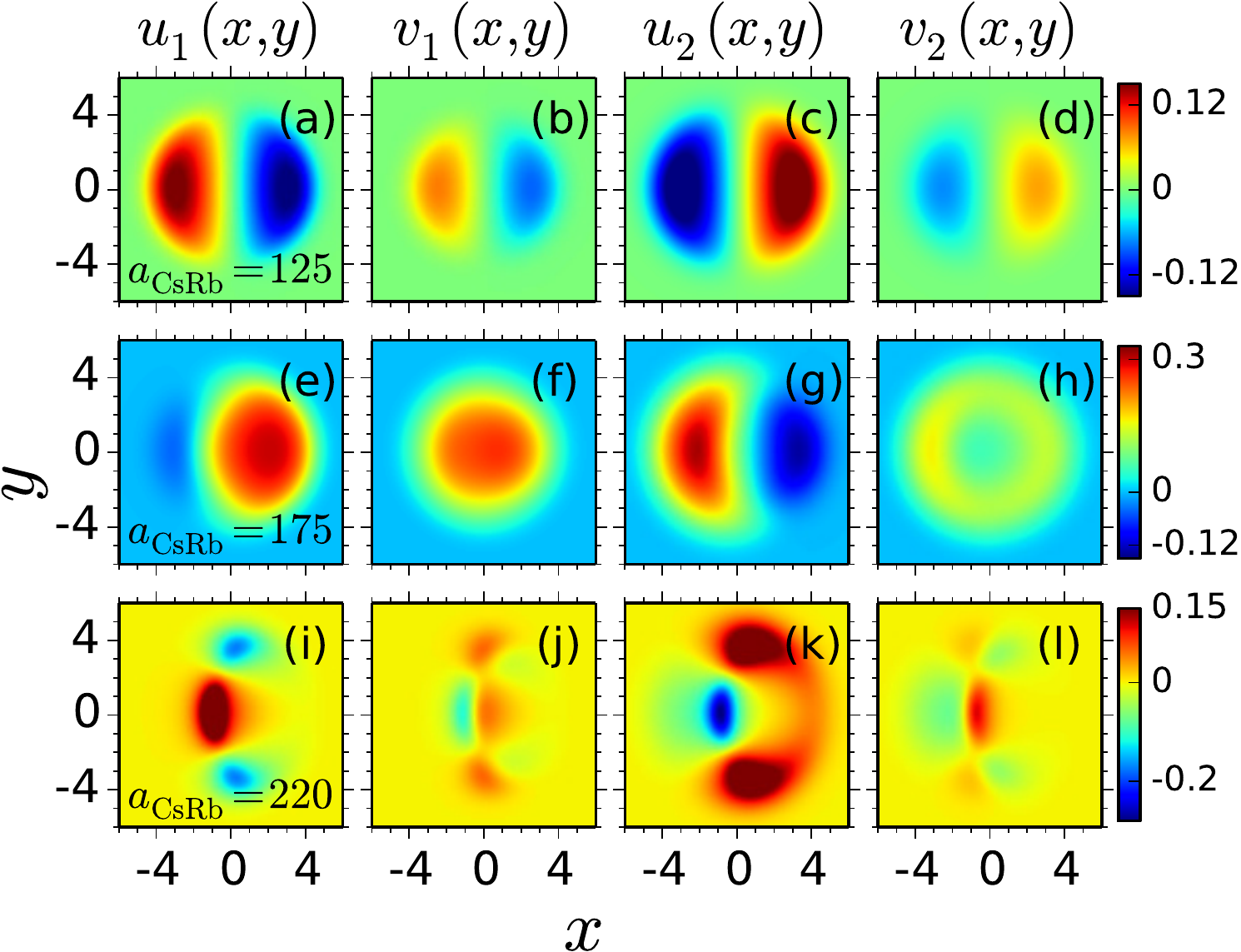}
      \caption{(Color online) Evolution of the quasiparticle amplitudes
                corresponding to the slosh modes whose energy
                decreases as $a_{\rm CsRb}$ is varied from
                $0a_0$ to $250a_0$. (a-d) show the quasiparticle amplitudes
                corresponding to one of the degenerate Kohn modes at
                $a_{\rm CsRb} = 125a_0$. At $a_{\rm CsRb} = 175a_0$, that is,
                at the point of phase-separation, the degeneracy of the Kohn
                mode is lifted and a bifurcation in the mode evolution is
                evident. The quasiparticle amplitudes corresponding
                to the upper branch of the slosh mode are shown in
                (e-h) for $a_{\rm CsRb} = 175a_0$, and 
                (i-l) for $a_{\rm CsRb} = 220a_0$.
                Here $u$s and $v$s are in units of
                $a_{\rm osc}^{-1}$. $x$ and $y$ are measured in units of
                $a_{\rm osc}$.
              }
      \label{hdqampCsRb}
\end{figure}

The transformation of the condensate density profiles leads to the 
modification of the structure of quasiparticle amplitudes as shown in 
Figs.~\ref{qampCsRb}, ~\ref{hdqampCsRb}. These amplitudes correspond to the 
slosh mode. We show that the metamorphosis of the quasiparticle 
amplitudes corresponding to the two branches are dramatically different as 
the degeneracy of the slosh mode gets lifted at the point of 
phase-separation. When $a_{\rm CsRb} = 125 a_0$, the condensates are 
miscible, and the slosh modes are degenerate with the 
same $|u_{\rm Cs}|$ and $|u_{\rm Rb}|$, but
are out-of-phase with each other. However, $|u_{\rm Cs(Rb)}|>|v_{\rm
Cs(Rb)}|$ as shown in Figs.~\ref{qampCsRb}(a)-(d), ~\ref{hdqampCsRb}(a)-(d).
At $a_{\rm CsRb}\approx 175 a_0$, the rotational symmetry of the ground
state is broken, and the condensates begin to phase-separate with 
the development of an interface. As mentioned earlier, the slosh mode 
bifurcates into two branches. The Bogoliubov quasiparticle amplitudes 
corresponding to the lower 
energy branch are shown in Fig.~\ref{qampCsRb}(e)-(h). The deformation of the
slosh modes with the breaking of rotational symmetry is evident. Here 
too $|u_{\rm Cs(Rb)}|>|v_{\rm Cs(Rb)}|$, but are out-of-phase with each other. 
For the amplitudes corresponding to the upper branch, the dipole nature of 
the modes begins to cease in $u_{\rm Cs}$ and $v_{\rm Cs}$, and becomes
almost single-lobed as shown in Fig.~\ref{hdqampCsRb}(e)-(f). Similar 
symmetry-broken structural deformation is demonstrated in $u_{\rm Rb}$ 
and $v_{\rm Rb}$ as shown in Fig.~\ref{hdqampCsRb}(g)-(h). After
phase-separation, that is when $a_{\rm CsRb}>175 a_0$, the non-degenerate
slosh modes represent both bulk and interface excitations. One of the slosh
modes belonging to the lower branch gets transformed to a Goldstone mode, and
the amplitude of this mode, at $a_{\rm CsRb}\approx 220 a_0$, resembling the 
condensate density profiles are as shown in Fig.~\ref{qampCsRb}(i)-(l). The 
upper branch, after phase-separation,
corresponds to the out-of-phase quasiparticle amplitudes describing the 
interface excitations which are localized along the interface separating the
condensates. These are shown in Fig.~\ref{hdqampCsRb}(i)-(l).

\subsubsection{Mode evolution for miscible to shell-structure transition}

\begin{figure}[t]
 \includegraphics[width=8.5cm]{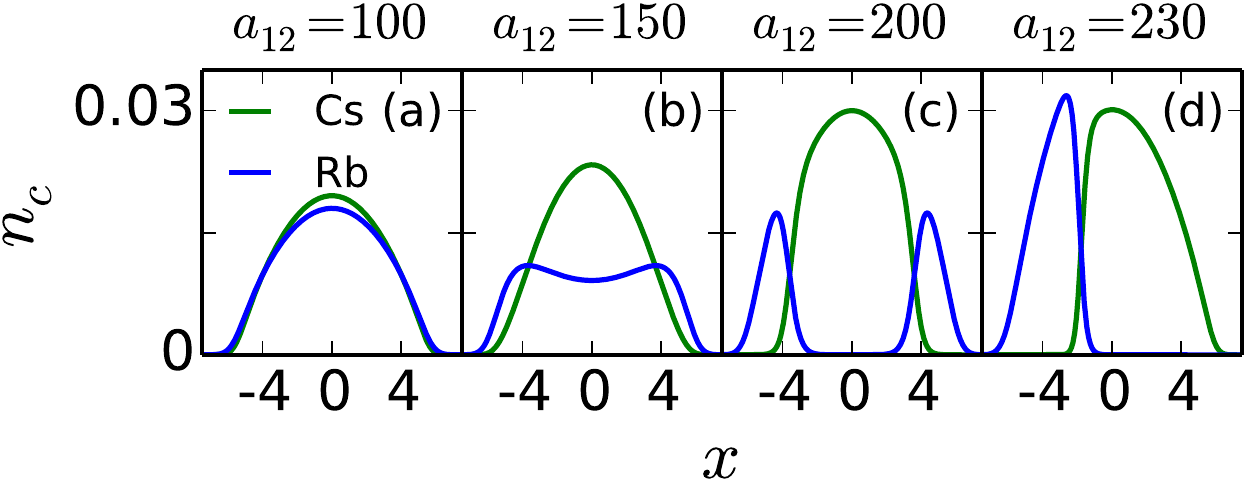}
       \caption{(Color online)Equilibrium density profiles of Cs-Rb TBEC
               along $x$-axis with $y = 0$ at $T = 0$ K showing transition
               from miscible to sandwich to side-by-side configuration with the
               change in interspecies scattering length $a_{\rm CsRb}$.
               (a, b) Shows Cs-Rb TBEC in for $a_{\rm CsRb} = 100 a_0, 150 a_0$
                      respectively.
               (c) Shows the phase-separated density profiles of Cs-Rb TBEC for
                   $a_{\rm CsRb} = 200 a_0$. These are referred to as
                   sandwich type density profiles.
               (d) Shows the phase-separated density profiles of Cs-Rb TBEC for
                   $a_{\rm CsRb} = 230 a_0$. These are referred to as
                   side-by-side density profiles. $n_c$ and $x$ are measured
                   in units of $a_{\rm osc}^{-2}$ and $a_{\rm osc}$,
                   respectively.
                }
    \label{sndCsRb}
\end{figure}

For higher number of atoms and higher $\omega_z$, the density profiles of the 
condensates acquire a different structure during the miscible-immiscible 
transition with variation in $a_{\rm CsRb}$. The structure is like a 
shell, where the Cs atoms occupy the center of the trap, and Rb atoms occupy 
the edges. This configuration emerges as the energetically favourable 
solution for intermediate values of $a_{\rm CsRb}$. However, for 
large $a_{\rm CsRb}$ the  
shell structured profiles get transformed to side-by-side density profiles. In 
which Cs atoms occupy one side of the trap, and Rb atoms occupy the other side. 
One typical example to demonstrate this trend in the evolution of density 
profiles is to consider $N_{\rm Cs} = N_{\rm Rb} = 5000$. The trapping 
parameters are $\omega_\perp = 2\pi \times 8.0$ Hz, 
and $\omega_z = 2\pi \times 200$ Hz which satisfy the quasi-2D condition, 
that is $\mu_k \ll \hbar\omega_z$.
For this set of parameters, when $a_{\rm CsRb} = 100a_0$, the ground state
of Cs-Rb mixture is in miscible phase. With increasing $a_{\rm CsRb}$, the
Rb condensate develops a dip at the center and becomes broader. At the
point of phase-separation, that is when $a_{\rm CsRb}\approx 200a_0$, 
shell structured density profiles emerge as the ground state with Cs atoms at 
the center surrounded by Rb atoms. Here, as mentioned earlier, the 
condensates assume a shell structure configuration for a narrow range 
of $200a_0\leqslant a_{\rm CsRb} \leqslant 205a_0$. That is to say, with a 
slight increase
in $a_{\rm CsRb}$ the side-by-side density profiles are the energetically
favourable ones. These variations in density distribution of the TBEC upon
increase in $a_{\rm CsRb}$ are shown in Fig.~\ref{sndCsRb}.

\begin{figure}[t]
 \includegraphics[width=9.0cm]{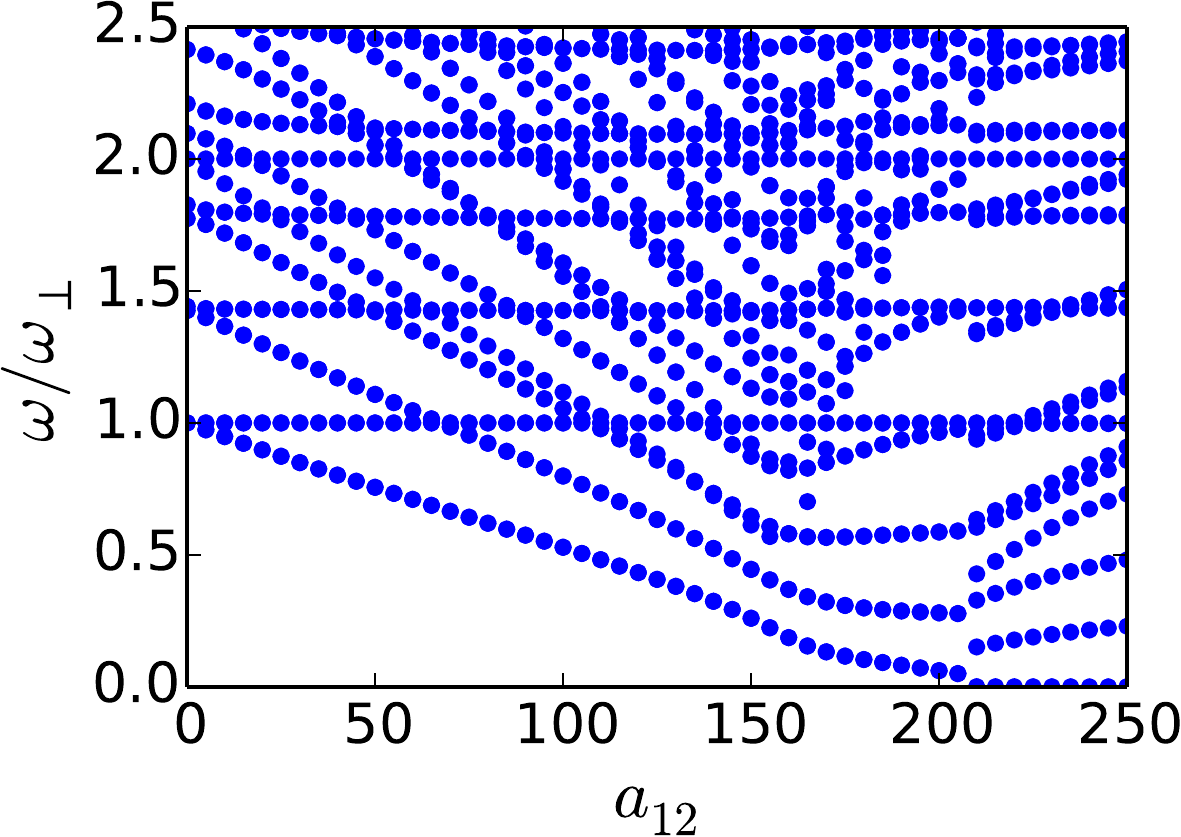}
     \caption{(Color online)The evolution of the low-lying modes in Cs-Rb TBEC
              corresponding to the transition shown in Fig.~\ref{sndCsRb}
              as a function of the interspecies scattering length
              $a_{\rm CsRb}$. Here $a_{12}$ is in units of $a_0$.
             }
    \label{sndwmevl}
\end{figure}

\begin{figure}[t]
 \includegraphics[width=8.5cm]{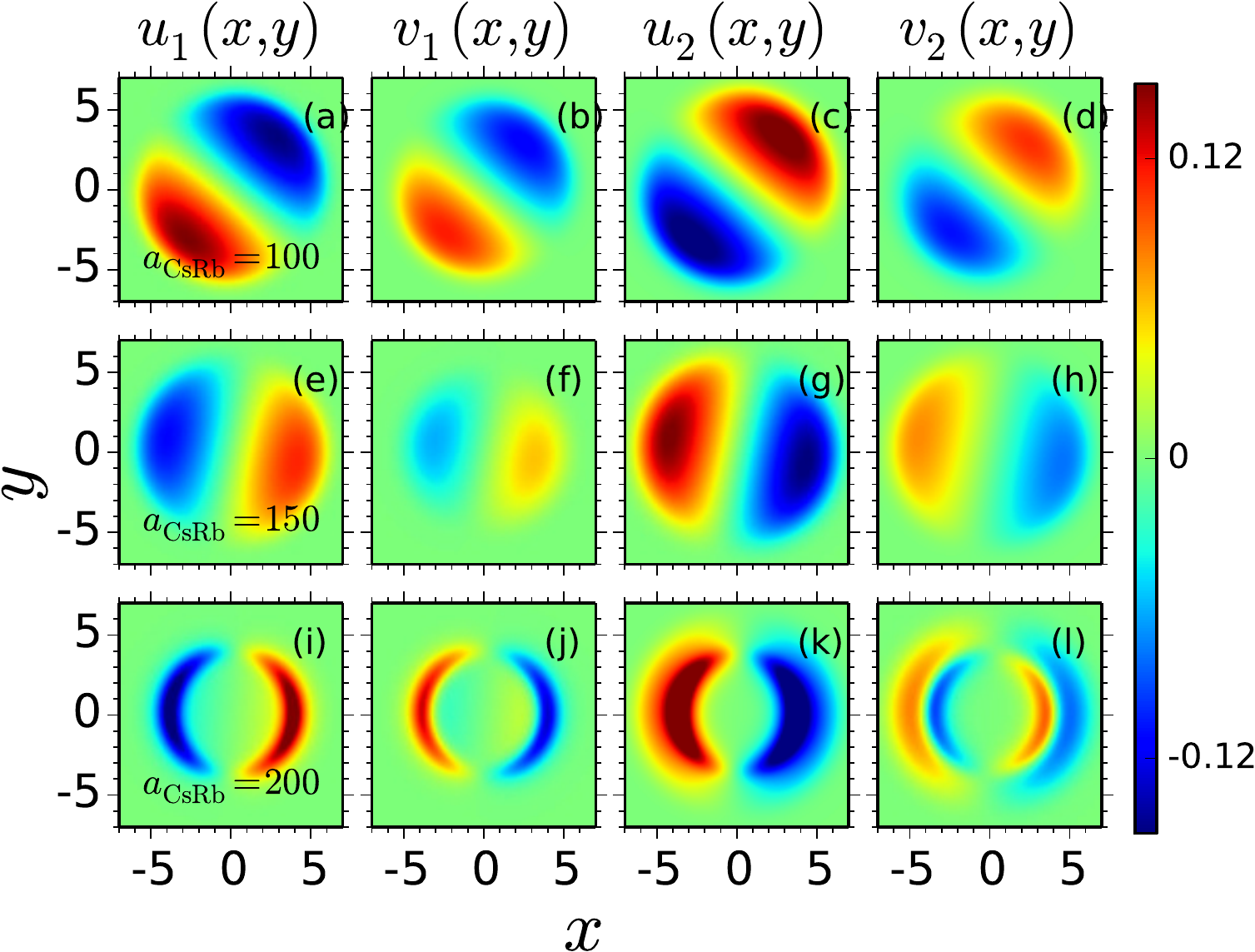}
    \caption{(Color online) Evolution of the quasiparticle amplitudes
                corresponding to the slosh modes whose energy
                decreases as $a_{\rm CsRb}$ is varied from
                $0a_0$ to $250a_0$. (a-d) show the quasiparticle amplitudes
                corresponding to one of the degenerate slosh modes at
                $a_{\rm CsRb} = 100a_0$. (e)-(h) show the quasiparticle
                amplitudes at $a_{\rm CsRb} = 150a_0$ when
                the condensates are partially miscible. (i)-(l) Shown here
                are the quasiparticle amplitudes corresponding
                to $a_{\rm CsRb} = 200a_0$. Here $u$s and $v$s are in units of
                $a_{\rm osc}^{-1}$. $x$ and $y$ are measured in units of
                $a_{\rm osc}$.
              }
    \label{sndqamp}
\end{figure}

\begin{figure}[H]
 \includegraphics[width=8.5cm]{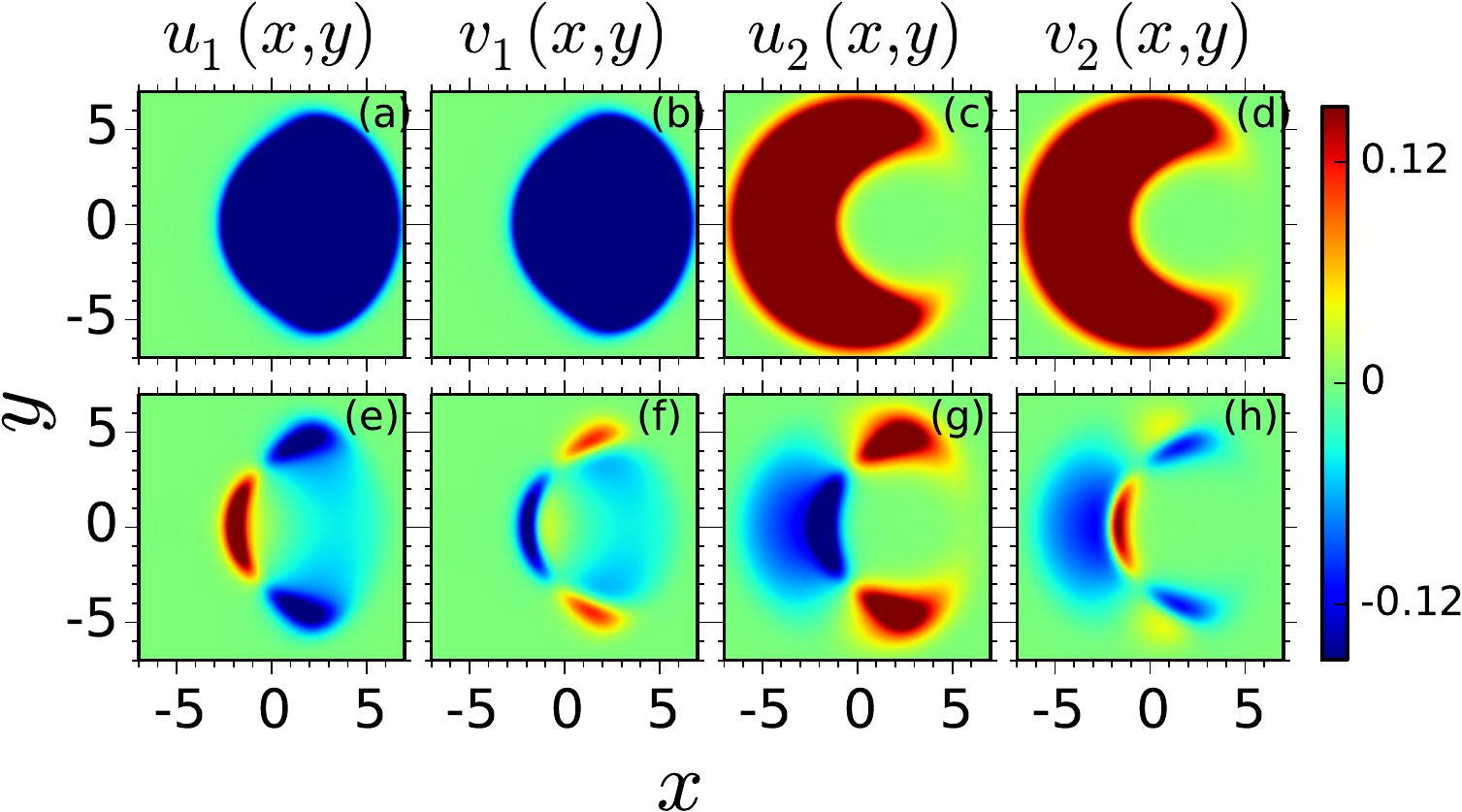}
    \caption{(Color online) Quasiparticle amplitudes corresponding to
             (a)-(d) the Goldstone mode, and (e)-(h) the interfacial
             excitations at $a_{\rm CsRb} = 230a_0$. Here $u$s and $v$s
             are in units of $a_{\rm osc}^{-1}$. $x$ and $y$ are measured
             in units of $a_{\rm osc}$.
            }
    \label{hsndqamp}
\end{figure}

With the change in $a_{\rm CsRb}$, like in the previous case, the energy of
the Kohn mode remains constant throughout the evolution. The energy of the
slosh mode, however, decreases. For $a_{\rm CsRb} > 200a_0$, the degeneracy of 
the slosh modes gets lifted giving rise to a bifurcation as shown in 
Fig.~\ref{sndwmevl}, and one of the slosh modes is transformed to
a Goldstone mode of the system. The energy of the other slosh mode hardens
indicating a symmetry breaking. This also reflected in the condensate density
profiles, as mentioned earlier, in this domain of $a_{\rm CsRb}$ the  
condensate density profiles has side-by-side geometry. Another indication in 
the excitation spectrum is the discontinuity as shown in 
Fig.~\ref{sndwmevl}.
Following this evolution, the Bogoliubov quasiparticle amplitudes
corresponding to the slosh mode undergo a significant change. When the
condensates are partially miscible, the slosh modes are degenerate
with same $|u_{\rm Cs}|$ and $|u_{\rm Rb}|$ but
are out-of-phase with each other. However, $|u_{\rm Cs(Rb)}|>|v_{\rm
Cs(Rb)}|$ as shown in Figs.~\ref{sndqamp}(a)-(d), (e)-(h) for two different
values of $a_{\rm CsRb}$. Furthermore, at $a_{\rm CsRb}\approx 200 a_0$, 
$|u_{\rm Cs}|<|u_{\rm Rb}|$ and are out-of-phase with each other, and the
slosh mode is deformed to an excitation along the axis of the
shell. These transformations of quasiparticle amplitudes are shown in
Fig.~\ref{sndqamp}(i)-(l). With a slight increase in $a_{\rm CsRb}$, when
the rotational symmetry is broken and as discussed earlier, the slosh mode 
bifurcates to become a Goldstone mode, and a higher excited mode representing 
the excitations along the interface of the condensate. For example, when 
$a_{\rm CsRb}\approx 230 a_0$, one of the slosh modes which have become the new 
Goldstone mode resembles the ground state density profiles as shown in 
Fig.~\ref{hsndqamp}(a)-(d). The other one corresponding to the out-of-phase 
interface excitations is shown in Fig.~\ref{hsndqamp}(e)-(h).

\section{Conclusions}
\label{conc}
The present studies reveal unique features in the nature of quasiparticle 
excitation spectrum of TBECs. In the immiscible domain, the position swapping 
of the constituent species in a $^{85}$Rb-$^{87}$Rb mixture, which is
driven by tuning intraspecies scattering length of $^{85}$Rb, is accompanied by 
the softening of slosh and quadrupole modes. These modes on getting softened
transform to zero energy modes at the point of equal intraspecies scattering 
lengths. These modes harden when the intraspecies scattering lengths $^{85}$Rb
increased to values higher than that of $^{87}$Rb.

For the Cs-Rb condensate mixture, where the atomic masses of the constituents
are widely different, we find a different trend in the mode evolution
spectrum. On steering the system from miscible to immiscible domain by
tuning the interspecies scattering length, the slosh mode softens. The
emergence of side-by-side density profiles as a result of phase-separation
is indicated by the breaking of the rotational symmetry, and the
bifurcation of degenerate slosh modes. The bifurcation gives rise to two
branches, of which, one becomes the Goldstone mode of the system.
Furthermore, the side-by-side density profile may also emanate from a
shell structured density profile indicated by a marked discontinuity in the
excitation spectrum. The variation in the quasiparticle excitations will
lead to differences in the non-condensate density distributions, and
dynamical structure factor which shall be investigated in our future works.


\begin{acknowledgments}
We thank K. Suthar, S. Bandyopadhyay and R. Bai for useful discussions.
The results presented in the paper are based on the computations using the
3TFLOP HPC Cluster at Physical Research Laboratory, Ahmedabad, India.
\end{acknowledgments}

\bibliography{mode}{}
\bibliographystyle{apsrev4-1}

\end{document}